\documentstyle[12pt,epsf]{article}
\renewcommand{\theequation}{\arabic{equation}}
\newcommand{\be}{\begin{equation}}
\newcommand{\ee}{\end{equation}}
\newcommand{\beqa}{\begin{eqnarray}}
\newcommand{\eeqa}{\end{eqnarray}}
\newcommand{\bean}{\begin{eqnarray*}}
\newcommand{\eean}{\end{eqnarray*}}
\newcommand{\eqn}[1]{(\ref{#1})}
\newcommand{\nn}{\nonumber}

\newcommand{\del}{\partial}
\newcommand{\gapproxeq}{\lower .7ex\hbox{$\;\stackrel{\textstyle >}{\sim}\;$}}
\newcommand{\lapproxeq}{\lower .7ex\hbox{$\;\stackrel{\textstyle <}{\sim}\;$}}
\def\up#1{\leavevmode \raise.16ex\hbox{#1}}
\newcommand{\journal}[4]{{ #1 }{ #2} \up(19#3\up) #4}
\newcommand{\snabla}{\nabla\!\!\!\!/}
\def\sqr#1#2{{\vcenter{\vbox{\hrule height.#2pt
	\hbox{\vrule width.#2pt height#1pt \kern#1pt
			\vrule width.#2pt}
	\hrule height.#2pt}}}}

\renewcommand{\theequation}{\thesection.\arabic{equation}}
\newcounter{appendice}
\newcommand{\appendice}
{
\setcounter{equation}{0}
\renewcommand{\theequation}{\Alph{appendice}.\arabic{equation}}
\addtocounter{appendice}{1}
{\bf Appendix \Alph{appendice}}
}

\def\thebibliography#1{\section*{REFERENCES\markboth
	{REFERENCES}{REFERENCES}}\list
	{[\arabic{enumi}]}{\settowidth\labelwidth{[#1]}\leftmargin\labelwidth
	\advance\leftmargin\labelsep
	\usecounter{enumi}}
	\def\newblock{\hskip .11em plus .33em minus -.07em}
	\sloppy
	\sfcode`\.=1000\relax}

\begin{document}
\setlength{\unitlength}{1mm}
\begin{flushright}
\small DFTUZ-98/32
\\ hep-th/9812076
\end{flushright}  

\vspace{1cm}
\begin{center}
{\large\bf Temperature induced phase transitions in four-fermion models 
in  curved space-time}
\end{center}
\bigskip\bigskip
\begin{center}
 {Patrizia Vitale}\footnote{on leave from 
{\it Dipartimento di Scienze Fisiche, Universit\`a di Napoli - 
Federico II - Mostra d'Oltremare Pad.  
19, I-80125, Napoli, Italy\\
$\scriptstyle{e-mail:~ vitale@na.infn.it}$
}}
\end{center}
\begin{center}
{\it Departamento de F\'{\i}sica Te\'orica. Facultad de Ciencias.\\
Universidad de Zaragoza, E-50009, Zaragoza, Spain.}
\end{center}
\vspace*{2cm}
\begin{abstract}
The large $N$ limit of the Gross--Neveu model is here studied
on manifolds with constant curvature, at zero and finite temperature. 
Using the $\zeta$--function regularization, the  phase
structure is investigated for arbitrary values of the coupling constant. 
 The critical surface 
where the second order phase transition takes place is analytically 
found for both the positive and 
negative curvature cases.
For negative curvature, where the symmetry is always broken at zero
temperature, the mass gap is calculated. 
The free energy density is evaluated at criticality and the zero
curvature and zero temperature limits are discussed.
\end{abstract}
\vspace*{1cm}
\noindent
\begin{center}
\end{center}
\vspace*{2cm}
\newpage
\baselineskip=.6cm

\section*{Introduction}
\setcounter{equation}{0}
The Gross--Neveu model \cite{GN} is the simplest version of models
with four-fermion interactions first introduced in \cite{NJL} as examples of
dynamical symmetry breaking. These models have been widely
studied in the literature in many different areas of theoretical physics:
for example 
as low energy effective theories of QCD \cite{muta}, or, in the Euclidean
formalism, as realistic
models describing 
phase transitions in statistical mechanics (see for example
\cite{jacobs}, where the stability conditions for the effective action in 
the large N limit are seen to imply the BCS gap equation of classical
superconductivity). In the cosmological framework
the massive composite field, $\sigma$, which arises after dynamical
symmetry breaking, has been proposed as an inflaton in inflationary
universe models \cite{mat}. With respect to each of the mentioned
problems
it is of interest to consider manifolds other than
$R^n$. In facts, the
compactification of one direction from $R$ to $S^1$ allows to describe 
$n-1$ dimensional systems at finite temperature. The introduction of scale
parameters in the other directions allows, in  low dimensional physics, 
 the study of the behaviour of dynamical systems under deformation
of their microscopic structure. Finally, the most obvious application 
of the study of this model in the  non-zero curvature case is the
cosmological framework.  

There exists a very reach
literature on four fermion models. An introduction
is contained in \cite{zinn1}. A recent and updated review 
may be found in \cite{IMO}. The
issue of 
critical 
exponents and $\beta$ function in flat space is addressed for example in
\cite{gracey,zinn2,kivel}. In \cite{KS}-\cite{EGOS} the 
effects of an external electromagnetic field in flat space-time are
considered. The  proof of $1/N$
renormalizability
in 3 dimensions was first addressed in 
 \cite{parisi,gross,rosen1}. The thermodynamical behaviour 
 in flat space-time is described in \cite{rosen2}-\cite{IKM}, 
 where the temperature induced second order phase transition is
found.
The influence of a classical
gravitational field on the dynamical symmetry breaking has been
analyzed in 
 \cite{BK}-\cite{KK}, where the existence of curvature induced phase
transitions is evidentiated. 
The relevance of  an external electromagnetic field to descriptions of  
the early universe is considered in a series of papers where the
electromagnetic effects are studied in combination with curvature 
effects \cite{GOS}-\cite{ES}.

In this paper we 
study the  large $N$ limit of the GN
model, 
on 3-d manifolds of constant, non-zero curvature, in the Euclidean
formalism. We focus our attention on the second order, temperature
induced, phase
transition and the r\^ole of curvature in the process of symmetry
restoration. This may be formalized by   considering the model on
manifolds of the form $\Sigma\times S^1$, where $\Sigma$ is a  
2-dimensional surface  with non-zero curvature and the inverse radius of the
circle plays the
r\^ole of the temperature (the compactification of one space-time
dimension to the circle is also referred as a finite size effect). To our
knowledge, the combined effect  of curvature and finite
temperature have only been considered in \cite{ELO} (in the weak curvature
approximation) and \cite{IMO}. In \cite{MV} the GN model is considered on
manifolds of the form $\Sigma\times S^1$, but the thermodynamical
phase transition is not discussed and the coupling constant is fixed to the 
flat space critical value. In non-perturbative
approaches it is a known result, although
yet debated, that positive
curvature and 
finite temperature enhance the process of symmetry restoration, while
negative curvature favours symmetry breaking.  
For
positive curvature, a finite temperature does not modify the qualitative 
features of the model but simply changes the phase transition point,
whereas, for negative curvature, there is no phase transition at $T=0$
(the symmetry is broken for any value of the coupling and of the
curvature), but as soon as the temperature is switched on a symmetry
restoration becomes possible, indeed it is realized, for some value of
$T$.

We consider the GN model 
in de Sitter and anti-de Sitter
backgrounds, respectively the manifolds 
$S^2\times S^1$ and $H^2\times S^1$. We apply  the large $N$
approximation, and use the zeta function regularization scheme. 
 The GN model exhibits
on $R^3$ a two--phase structure, the phase transition occurring for a
non--trivial  value of the coupling constant, the ultraviolet (UV)
fixed point. 
When considering the model in curved backgrounds, at finite temperature, 
the Lagrangian turns out to be  dependent
on three parameters: the coupling constant, $q$,  the parameter
$\beta$ (the inverse temperature), and  the curvature
parameter, $r$.
Phase transitions can in principle occur with respect to any of  them.
With respect to the
thermodynamical characteristics,  we recover the 
qualitative behaviour of \cite{IMO}, and we find a simple 
analytic  expression for the critical  surfaces $f(r,\beta,q) = 0$. 
 We follow  the conventions of \cite{GRV} and \cite{MV}  
 where the zeta function regularization is adopted to
study the non-linear sigma model and the GN model, respectively,  at the
critical value of the coupling constant.
 We find some interesting
common features between  the GN model and the non-linear $\sigma$ model.
For example, the mass gap on $H^2\times R$ at the critical coupling, for
the GN model is
found to be half the mass gap of the non-linear $\sigma$ model on the same
manifold. 

In section 1 we briefly review the properties of the
GN model and the large $N$ limit. 
Then we  recall the flat space
analysis in the zeta-function
regularization.   The curvature and temperature
effects are considered in section 2.  The two appendices are devoted to  
calculations related to section 2. 

\section{The Gross--Neveu model in three dimensions}
\setcounter{equation}{0}
In this section, upon giving some basic definitions for the study of
the model on a generic Riemannian manifold in the $1/N$ expansion, we
review the zeta function regularization scheme as discussed in
\cite{MV}. Then, to fix the notation and to illustrate the method in 
a simple case, we analyze the model in the Euclidean flat space
$R^3$. 

The Gross--Neveu model
on a Riemannian manifold $({\cal M},g)$  is described in terms of a $O(n)$
symmetric action for a set of $N$ massless Dirac fermions. The
Euclidean partition function  in 3--dimensions in the
presence of a background metric $g_{\mu \nu}(x)$ is given by 
\be 
{\cal Z}[g]=\int {\cal D}[\psi]~ {\cal D}[{\bar\psi}]~ 
\exp\left\{\int_{\cal M} d^3x~
\sqrt{g}~\left[{\bar \psi}_{i}(x) \snabla \psi_i(x) 
+ {q\over 2} ({\bar\psi}_{ i} \psi_i)^2\right]\right\}~~~,
\label{genfun}
\ee
where $i=1,2,\cdots,N$, $\snabla$ is the Dirac operator
on ${\cal M}$, and $q$ is the coupling constant\footnote{According to our
notation the Dirac matrices obey the following algebra: $\gamma_{\mu}
=  \gamma_{\mu}^{\dag}$, $\left\{ \gamma_{\mu}, \gamma_{\nu} \right\}
= 2 g_{\mu \nu}$, and $\mbox{Tr}\left(\gamma_{\mu}\right) = 0$.
Thus, the Dirac operator is antihermitian $\snabla^{\dag} = - 
\snabla$.}. 

The Dirac matrices are given in
terms of the Pauli matrices $\sigma_a$ by the expression 
\be
\gamma_\mu=V_{\mu,a} \sigma_a~~~,~~~~~\mbox{with}~~~\mu,a=1,2,3
\label{II.2}
\ee
where $V_{\mu,a}$ denote the {\it dreibein} defined by the equation
\be
g_{\mu\nu}=V_{\mu,a}(x) V_{\nu,b}(x) \delta_{ab}~~~. 
\label{II.3}
\ee
The {\it covariant derivative} $\nabla_{\mu}$ acting on a spinor field
is defined as \cite{zinn1}, 
\be
\nabla_{\mu} = \partial_{\mu}
+ \Gamma_{\mu}(x)~~~,
\label{II.4}
\ee
where $\Gamma_{\mu}$ is the spin connection
\be
\Gamma_{\mu}(x) \equiv  { 1 \over 8} \left[ \sigma_{a},\sigma_{b}
\right] V^{\nu}_{a}\left( \nabla_{\mu} V_{\nu,b}\right)~~~.
\label{II.5}
\ee
In even dimensions the model has a discrete chiral symmetry which 
prevents the addition of a fermion mass term, while in odd dimensions a 
mass term breaks space parity.

As it is usually done, the partition function \eqn{genfun} is rewritten
by introducing an 
auxiliary scalar field $\sigma$, such that
\be
{\cal Z}[g]=\int {\cal D}[\psi]~{\cal D}[{\bar\psi}]~{\cal D}[\sigma] 
\exp\left\{\int_{\cal M} d^3 x~\sqrt{g}~ 
\left[{\bar\psi}_{ i} (\snabla + \sigma) \psi_i 
-{1\over 2 q} \sigma^2\right]
~ \right\}~~~.
\label{gfun}
\ee
The $\sigma$ field has no dynamical effect, its introduction amounting to
multiply the partition function by an overall constant. 
We note here that in the large $N$ approximation the generating functional
\eqn{gfun} describes the Nambu-Jona-Lasinio model as well, when the
space-time dimension is greater than 2. This model differs by the GN model
in the presence of an extra term in the action. It reads
\be
S_{NJL}= S_{GN} -\int_{\cal M} d^3 x~\sqrt{g}~(\bar\psi_i\gamma_5\psi_i)^2.
\ee
Following the same method as above, it can be seen that the new
interaction gives rise to another auxiliary field, called $\pi$ in the
literature. In the large $N$ limit, for $d> 2$, $\pi$ can be reabsorbed
into the definition of the $\sigma$ field, giving rise to the generating
functional \eqn{gfun} (see for example \cite{IMO} for details).

 The generating functional $\cal
Z$ has to be regularized in the ultraviolet. This is usually done  by   
introducing a cut-off, $\Lambda$, in the momentum space. 
By redefining  the dimensional coupling constant $1/ q(\Lambda)$
as $\Lambda/ q'(\Lambda)$, the regularized
partition function is formally rewritten as 
\begin{eqnarray} 
{\cal Z}[g, \Lambda ]=\int {\cal D}_{\Lambda}[\psi]~
{\cal D}_{\Lambda} [{\bar\psi}]~
{\cal D}_{\Lambda}[\sigma]~
\exp\left\{\int_{\cal M} d^3x~ \sqrt{g}~ 
\left[{\bar\psi}_{ i} (\snabla + \sigma) \psi_i -{\Lambda\over 2 q'}
\sigma^2 
\right]\right\},
\label{II.11}\nonumber\\
\end{eqnarray}
where ${\cal D}_{\Lambda}[\psi]=\prod\limits_{|k|<{\Lambda}} d\psi (k)$ and
similarly for the other fields (we will write $q$ without the prime from now
on).

As for the non linear $\sigma$ model, the
existence of a non trivial UV fixed point shows that the large
momentum behaviour is not given by perturbation theory above 2
dimensions, where the theory is asymptotically free (see for
example\cite{zinn1,IMO}). We use here 
 the $1/N$ expansion, which has the  property of providing a
renormalizable theory for the 3-d GN model
 \cite{parisi, gross, rosen1}. 
	
In the large $N$ limit, which means $N \rightarrow \infty$ keeping $N
{q(\Lambda)}$ fixed, the generating functional can be calculated using
the saddle point approximation. For this purpose we integrate over
$N-1$ fermion fields, rescale the remaining fields $\psi_{N}$,
${\bar\psi}_{N}$ to $\sqrt{N-1}~~ \psi_{N}$, $\sqrt{N-1}
~~{\bar\psi}_{N}$, respectively, and redefine $(N-1) {q(\Lambda)}$ as
${q(\Lambda)}$. Thus we get 
\begin{eqnarray} 
{\cal Z}[g,\Lambda,q(\Lambda)]=\int {\cal D}_{\Lambda}[\psi_{N}]~ 
{\cal D}_{\Lambda}[{\bar\psi}_{N}]~{\cal D}_{\Lambda}[\sigma]~
\exp\left\{ (N-1)\mbox{Tr}\log_{\Lambda} (\snabla + \sigma)
\right\} \nonumber\\
\times \exp\left\{ (N-1)\int_{\cal M} d^3x~ \sqrt{g}~
\left[{\bar\psi}_N(\snabla +\sigma) \psi_N
-{\Lambda \over {2 q}}\sigma^2(x)\right]\right\}~~~.
\label{parfun}
\end{eqnarray}
In the limit $N \rightarrow \infty$ the dominating contribution to the
functional integral comes from the extremals of the action. For an
arbitrary metric $g_{\mu \nu}(x)$, these are obtained by extremizing
the action with respect to $\psi_{N}(x)$ keeping $\sigma(x)$ and
${\bar\psi}_N (x)$ fixed and vice--versa. Hence, a set of equations
({\it gap equations}) is obtained 
\beqa
{\bar\psi}_{N} (\stackrel{\leftarrow}{\snabla} -\sigma)=0~~~,
\label{gap1} \\
(\snabla +\sigma) \psi_N=0 ~~~,\label{gap11}  \\
G_{\Lambda}(x, x; \sigma,g)+ {\bar\psi}_N \psi_N - {\Lambda \over
{q(\Lambda)}} \sigma =0 ~~~, 
\label{gap2}
\eeqa
where $G_{\Lambda}(x, x; \sigma,g)\equiv\langle x|(\snabla + \sigma)^{-1}|x
\rangle_{\Lambda}$ is the two-points correlation function of the
$\psi$-field, evaluated for $x \rightarrow x'$. 
At the saddle point Eq. \eqn{parfun} reads
\begin{eqnarray} 
{\cal Z}[g,\Lambda,q(\Lambda)]=
\exp\left\{ (N-1)\left[\mbox{Tr}\log_{\Lambda} (\snabla + \sigma)
- {\Lambda \over {2 q}}\int_{\cal M} d^3x~ \sqrt{g}~
\sigma_c^2(x)\right]\right\}~~~.
\label{zzero}
\end{eqnarray}	
Then, the free energy density, 
\be
{\cal W}[g,\Lambda, q(\Lambda)]= {\log{\cal Z}\over N \int
d^3 x \sqrt g }\label{fren},
 \ee
may be calculated.  
In the following analysis we will look for uniform solutions of the gap 
equations 
\be
\langle \sigma \rangle = m~~~,~~~ 
\langle \psi_N \rangle = b~~~,~~~
\langle {\bar\psi}_N \rangle={\bar b}~~.
\label{gap}
\ee
 The quantities $b$ and ${\bar b}$ represent the
vacuum expectation value (v.e.v.) of the fermion fields, while $m$, if
positive, can be regarded as the mass of the field fluctuations around
the vacuum. 

Substituting these values into Eq. \eqn{fren} we have
\be 
{\cal W}[m, g, \Lambda,q(\Lambda)]
= \left[\log_\Lambda\det(\snabla +m) -{\Lambda \over 2q(\Lambda)}
~{m^2} \right]~~~. 
\label{freen}
\ee
Phase transitions occur at the extrema of the effective potential,
$V(\sigma)$, which is defined in the usual way: introducing a source $J$
for the $\sigma$ field in the generating functional \eqn{gfun}, we have
\be
 {\cal Z}[J]=\int {\cal D}[\psi]~{\cal D}[{\bar\psi}]~{\cal D}[\sigma] 
\exp\left\{-S(\psi,\bar\psi,\sigma)+ J\sigma ~ \right\}~~~;
\ee
we then introduce the Legendre transform
\be
\Gamma[\tilde\sigma]=-\log {\cal Z}[J]+ J\tilde\sigma
\ee
where $\tilde\sigma={\delta\log{\cal Z}[J]\over\delta J(x)}$.
The effective potential is defined to be
\be
V(\tilde\sigma)={\Gamma[\tilde\sigma]\over\int d^3 x \sqrt{g} }.
\ee
When evaluating the partition function in  the large $N$ limit it is found
that $\tilde\sigma=m$ and the effective potential is equal to 
{\it minus} the free energy density \cite{zinn1}. 
Then, the gap equations, \eqn{gap1}-\eqn{gap2} are the extrema of the
effective potential as well.

Instead than using the cutoff regularization introduced above, we use the
zeta-function regularization, which seems to us easier to handle in the
presence of curvature. 
Following  \cite{MV}, the equal point
Green's function $G(x,x,\Lambda)$ for
the operator $\snabla +m$ is seen to be regularized as 
\be
G_s(x,x;m,g)= m\langle x|(\Delta_{1/2}+m^2)^{-s}|x
\rangle=m\; \zeta(s,x)~~~,
\label{gref}
\ee
so that 
\be 
G(x,x;m,g)=m~\lim_{s \rightarrow 1} \zeta(s,x)~~~.
\label{grefu}
\ee
 $\zeta(s,x)$ is the local zeta function
\be
\zeta(s,x)=\sum_{n} (\lambda^2_n+m^2)^{-s} |\psi_n(x)|^2~~~,
\ee
\label{zetaf}
 $\lambda^2_{n}+m^2$ are the eigenvalues of the operator
$(\Delta_{1/2}
+ m^2)$ and $\left\{ \psi_{n}(x)\right\}$ is an orthonormal basis of
eigenvectors ($\Delta_{1/2}$ is the squared Dirac operator). 
 The sum over the eigenvalues includes degeneracy and in case of a
continuous spectrum the sum is replaced by an integral. 
 On homogeneous spaces such as the ones we
will be considering in this paper,
$\zeta(s,x)$ turns out to be independent of $x$. Moreover the equal point
Green's function happens to be finite for the 3-d case, as will be clear
in all the situations considered below (this is not the case in $d\ne 3$).
 
The gap equations \eqn{gap1}-\eqn{gap2} become, in this regularization
scheme
\beqa
\bar b(\gamma^\mu \Gamma_\mu -m)&=& 0 \label{ggap1}\\
(\gamma^\mu \Gamma_\mu +m)b&=& 0 \label{ggap11}\\
 m \lim_{s\rightarrow 1}\biggl\{ \frac{1}{q(s)} 
-\zeta(s,m)\biggr\} - {\bar b} b &=& 0 ~~~,
\label{ggap2}
\eeqa
where the regularized coupling $\Lambda/q(\Lambda) $ has been replaced by 
$1/q(s)$.
The free energy density \eqn{freen} can be written in terms of the
zeta function in its turn, recalling  that
\be
\log\det(\snabla +m) = {1\over 2} \log\det(\Delta_{1/2} +m^2)=-{1\over
2}\biggl[{d\over ds}\zeta(s,m)\biggr]_{s=0}.
\ee
We have then
\be 
{\cal W}[g,q,m]
= -{1\over 2}\biggl[{d\over ds}\zeta(s,m)\biggr]_{s=0} 
-{1\over 2q}~{m^2} ~~~, 
\label{regfen}
\ee
where ${1/ q}= \lim_{s\rightarrow 1} 1/q(s)$ is the renormalized coupling,
defined by Eq. \eqn{ggap2}.

Let us see how this scheme applies to the case of $R^3$. 
 Upon substituting the appropriate 
eigenvalues of the Dirac
operator for this space, the gap equations \eqn{ggap1}-\eqn{ggap2} now 
read
\beqa m\bar b= m b = 0 \\
m \lim_{s\rightarrow 1}\biggl\{ \frac{1}{q(s)} 
-\int {d^3k\over (2 \pi)^3}{1\over (k^2 +m^2)^s} \biggr\} - {\bar b} b &=&
0 ~~~.
\eeqa
Using the Mellin transform to analytically continue the zeta function, 
\be
\zeta(s)={1\over \Gamma(s)} \int dt~t^{s-1} h_{R^3} (t; x=x') ´e^{-m^2 t}
\label{mellin}
\ee
where $h_{R^3} (t; x=x') = (4\pi t )^{-3/2}$ is the equal point
heat kernel of the spin-${1\over 2}$ Laplacian, 
and
observing that  $b$ is zero, be $m$ zero or not, we are left with
\be
m \lim_{s\rightarrow 1} \biggl({1 \over q (s)}
- {m^{3-2s} \over {(4 \pi)}^{ 3 \over 2}}
{\Gamma(s-{3 \over 2})\over {\Gamma (s)}}\biggr)=0 . \label{firder}
\ee
Performing the limit we observe that 
$m=0$ for positive values of the coupling, while it may be non zero for
negative values of the coupling. At  $m=0$ the second derivative of
the effective potential changes its sign, being positive when
$1/q<0$, negative when $1/q>0$ (we recall that Eq. \eqn{firder} is 
the first derivative of the effective
potential, up to a minus sign). Thus, the system undergoes a second
order phase transition and the critical value of the coupling is 
\be
{1\over q_{cr}} =0~. \label{qcr}
\ee
The mass gap in the broken phase is given by
\be
m_0=-{4\pi\over q} \label{m0gap}
\ee
where the index zero will serve as future reference to distinguish 
 flat space quantities. Recalling that the
critical value of the coupling is independent on  the curvature but 
dependent on the regularization,
the result \eqn{qcr}
will be valid, in our regularization scheme, for all the manifolds
considered, while Eq. \eqn{m0gap} will be taken as the definition of the
renormalized coupling when it is negative.
From Eq. \eqn{regfen} the free energy density is now easily calculated to be  
\be
{\cal W}= -{m^2\over 2}\biggl({m\over 6\pi}+{1\over q}\biggr)
\label{fren0}
\ee
which is seen to be zero for $1/q\ge 0$ (m being zero), positive for $1/q$
negative;
in the latter case we have
\be
{\cal W}_0= {m^3\over 24\pi}. \label{freen0}
\ee

\section{The Gross-Neveu model in curved space-time}
\setcounter{equation}{0}
In this section we study the large $N$ limit of the Gross--Neveu model
on manifolds of the type $\Sigma\times S^1$ where $\Sigma$ is a two
dimensional manifold of constant curvature. The $S^1$ component can be
regarded either as
a compact space-time dimension, or as a way of introducing the temperature
through 
the inverse radius of the circle. The former point of view is 
adopted when studying finite size effects. Here we will adopt the latter 
point of view. 
 
\subsection{The manifold $S^2 \times S^1$}
This is a positive curvature space-time, of  scalar curvature
${\cal R} = {2\over r^2}$ ($r$ is the radius of the sphere).
As we already mentioned in the introduction, it is expected that both the
positive curvature and the finite temperature favour the symmetry
restoration.
This result is confirmed in our approach, where
we find a second order phase transition at some finite value of $T=f(r)$. 
When either the curvature or the temperature tend to zero, the two-phase
structure persists recovering  some known results.

 We parameterize this space by $x^{\mu} \equiv (\tau,~
\chi,~\theta)$, 
where $0\le\tau< 2\pi$, $-\pi/2\le\chi\le \pi/2 $,
and  $0\le\theta< 2\pi$. The metric tensor is then defined as 
\be
g_{\mu \nu} dx^\mu \otimes dx^\nu=r^2~\cos^2\chi~d\tau \otimes d\tau
+ r^2 d\chi \otimes d\chi+ \beta^2 d\theta \otimes d\theta~~~. 
\label{met}
\ee
The Dirac matrices  are given in terms of the Pauli matrices $\sigma_a$,  by  
\be
\gamma_{1} = r \cos\chi ~ \sigma_1~~~,~~~~~~~~~
\gamma_{2} = r \sigma_2~~~,~~~~~~~~\gamma_3=\\ \beta \sigma_3
~~~.
\label{dima}
\ee
The spin connection \eqn{II.5} results to be 
\be
\Gamma_{\mu}(x) =
- {i \over 2} \sigma_{3} \sin{\chi}~\delta_{1\mu}~~~.
\label{spco}
\ee
Due to the form of the spin connection, it was seen in \cite{MV}
that the vacuum expectation value of
the $\psi$ fields has to be zero, for the first two gap equations to be
satisfied.  Hence we are left with the gap
equation
\be
 m \lim_{s\rightarrow 1}\biggl\{ \frac{1}{q(s)} -\zeta(s,m) \biggr\} = 0
~~~. \label{gaps2s1}
\ee
Observing that the heat kernel of the Laplacian on a product manifold is
just the product of the heat kernels on the factor spaces, and observing
that the equal points heat kernel Laplacians on $S^2$ and $S^1$ are
respectively
\beqa
h_{S^2}(t)&=&{1 \over 2\pi r^{2}}\sum_1^{\infty} l \exp\left(-{l^2\over
r^2} t\right) \label{hkers2}\\
 h_{S^1}(t)&=&{1 \over \beta}\sum_{-\infty}^{\infty} 
\exp\left[-4 \pi\left(n+ {1\over
2}\right)^2 t\right] \label{hkers1}
\eeqa
the zeta-function is obtained by Eq. \eqn{mellin} to be 
\be
\zeta(s,m)= 
{\beta^{2s-1}\over 2 \pi^2 r^2
\Gamma(s)} \int^{\infty}_0 dt~ t^{s-1}
\sum_{n=-\infty}^{\infty} \sum_{l=1}^{\infty} l \exp \Biggl\{-
\left[4\pi^2 \left( n+{1 \over 2}\right)^2  + l^2{\beta^2 \over r^2}
+m^2\beta^2 \right]t\Biggr\}.
 \label{zs2s1}
\ee
Upon evaluating the integral and taking  the limit $s\rightarrow 1$ the
gap equation \eqn{gaps2s1} reads
\beqa
{m\over q}&=& 
 {m^{2}  \over 2 \pi^{2}} 
\int^{\infty}_0 {\mbox P}\int_{0}^{\infty} dx ~ \left(\cot(x)
-{1\over x} \right) K_{1}(2mxr) \nn\\
&-& {m^2 \over 4\pi} +
{m\over2\pi r} 
\sum_{n,l=1}^{\infty} (-1)^n {l\over\sqrt{l^2+m^2 r^2}} 
\exp \left(-n{\beta\over r} \sqrt{l^2 +m^2 r^2}\right) ,\label{lgaps2s1}
\eeqa
where P indicates the principal value of the integral. The details of the
calculation are given in  appendix A.
Let us analyze the possible solutions. We already found in \cite{MV}
that $m=0$ is the only solution at the critical coupling $1/q_{cr}=0$.
In facts, $m$ is equal to zero for $1/q\ge 0$, the RHS of
Eq. \eqn{lgaps2s1} being negative
for any value of $\beta$, $r$.  Due to this, $m$ can
be non zero only for negative values of the coupling. Let us fix 
$1/q<0$.
On deriving Eq. \eqn{lgaps2s1} with respect to $m$, it may be seen that, at
$m= 0$ the second derivative of the effective potential changes its sign
when varying $\beta,r$ around some $\beta_{cr}, r_{cr}$. This is where a
second order phase transition occurs \cite{IMO}. We propose here a
simple expression for the critical surface $f(\beta, r, q)=0$.
To find the critical surface we use the fact that in a second order
phase transition the mass gap smoothly disappears as approaching
criticality. Thus, we perform the limit $m\rightarrow 0$ for the
equation
\beqa
-{m_0\over 4\pi}&=& 
 {m  \over 2 \pi^{2}} 
\int^{\infty}_0 {\mbox P}\int_{0}^{\infty} dx ~ \left(\cot(x)
-{1\over x} \right) K_{1}(2mxr) \nn\\
&-& { m \over 4\pi} +
{1\over2\pi r} 
\sum_{n,l=1}^{\infty} (-1)^n {l\over\sqrt{l^2+m^2 r^2}} 
\exp \left(-n{\beta\over r} \sqrt{l^2 +m^2 r^2}\right) ,\label{limgaps2s1}
\eeqa
which is obtained by Eq. \eqn{lgaps2s1} by factorizing out a common $m$ and
replacing the negative coupling with its renormalized value \eqn{m0gap}.
Recalling that $K_\nu(x)\stackrel{x\rightarrow 0}{\rightarrow} 2^{\nu-1}
(\nu -1)! x^{-\nu}$, we obtain
\beqa
-m_0&=& 
 {1  \over  \pi r} 
\int^{\infty}_0 {\mbox P}\int_{0}^{\infty} {dx\over x} ~ \left(\cot(x)
-{1\over x} \right)  
+ {2\over r} \sum_{n,l=1}^{\infty} (-1)^n  
\exp \left(-n{\beta\over r} l\right)\nn\\
&-&{1\over 2r} + {2\over r} 
\sum_{n=1}^{\infty} {(-1)^n\over   
e^{n{\beta\over r} } -1} .
\eeqa
The principal value of the integral may be performed, yielding 
\be 
r'_{cr}= {1\over 2}- 2 \sum_1^{\infty} {(-1)^n\over   
e^{n{\beta'_{cr}\over r'_{cr}} } -1}, \label{s2s1crili}
\ee 
where the critical radii have been rescaled by the flat space mass gap,
 $m_0$. The series is convergent, indeed its sum is a hypergeometrical
function. 
We recall that this expression is only valid for negative
values of the coupling.
In Fig. 1 we plot the critical
parameter $1/ \beta'_{cr}$ versus the  critical curvature parameter $1/
r'_{cr}$.
The relation \eqn{s2s1crili}  is particularly easy to handle when
the
zero curvature limit and the zero temperature limit are to be performed.
For $r\rightarrow \infty$ we obtain the known result \cite{rosen2}
\be
\left({1\over \beta_{cr}}\right)_{R^ 2\times S^1}={m_0\over 2 \log 2}.
\label{0curv}
\ee
For $\beta\rightarrow \infty$ we get
\be
\left(r_{cr}\right)_{S^2\times R}={1\over 2 m_0}, \label{0T}
\ee
reproducing the result of \cite{IIM}.  

To evaluate the free energy density, defined in Eq. \eqn{regfen}, we need to
calculate the derivative of the zeta-function \eqn{zs2s1}. This is
done in appendix A in some detail. The result is 
\beqa
{\cal W}(m) &=& -{m^2 \over 2 q} -
{m^{2} \over 4 \pi^{2} r } 
\int^{\infty}_0 {\mbox P}\int_{0}^{\infty} dx ~{1\over x} \left(\cot(x)
-{1\over x} + {x\over 3}\right) K_{2}(2mxr) \nn\\
&-& {m^{3} \over 12 \pi}
-{m\over 48 \pi r^2}  
- {1\over 2\pi\beta r^2}
\sum_{n,l=1}^{\infty} {(-1)^n\over n} l
\exp\left[-n{\beta\over r} \sqrt{l^2\ + m^2 r^2}\right] \label{s2s1fe}
\eeqa
 Different limits can be performed easily (zero temperature and/or
zero curvature, zero mass), reproducing some known results. In the
limit of zero curvature and temperature we recover the flat space limit
\eqn{fren0}.  Here we just
consider the zero mass limit, which allows us to calculate the free energy 
at criticality. For $m\rightarrow 0$, Eq. \eqn{s2s1fe} simplifies
considerably. We obtain
\be
{\cal W}_{S^2\times S^1}(m=0)= -{1\over 8 \pi \beta r^2}\sum_{n=1}^{\infty} 
{(-1)^n\over n} \mbox{cosech}^2({\beta\over 2 r} n).
\ee
This result, already found in \cite{MV}, gives the free energy at the
critical line, when $\beta$ and $r$ satisfy Eq. \eqn{s2s1crili}. 
Moreover, it reproduces the correct result at zero curvature
\cite{klime,MV}
\be
{\cal W}(m=0)_{R^2\times S^1}= {3\over 8\pi \beta^3} z(3)
\label{0cu}
\ee
where $z$ is the Riemann zeta function.

\subsection{ The manifold $H^2 \times S^1$}
We now consider the product manifold $H^2_r
\times S^1_\beta$,
where 
$H^2_r$ is the 
2--dimensional {\it pseudosphere}. This is a space with constant negative
curvature. Previous calculations suggest that the negative curvature
favours symmetry breaking so that we have to expect a competitive
effect between curvature and temperature \cite{IMO}. This behaviour is
indeed
confirmed in our analysis. We find a second order phase transition for
some finite $T=f(r)$, while the symmetry is always broken at zero $T$. 
The latter case, which may be formalized by considering the model on 
the manifold $H^2\times
R$, is studied in some detail and the mass gap is evaluated.

We parameterize $H^2_r$ as
$~~~
H^2_r=\{ z = (x, y),~~ x\in R,~~ 0 <y<\infty \}~~~,$
while the circle $S^1_\beta$  is parameterized as before
by $\theta$, $0 \leq \theta <2\pi$. The scalar curvature of $H^2_r$
is ${\cal R} = -2/r^2$, where $r$ is a constant positive parameter.
The metric tensor on the whole manifold is then given by
\be
g_{\mu \nu} dx^\mu \otimes dx^\nu
={r^2 \over y^2} \biggl(dx\otimes dx + dy \otimes dy
\biggr) + \beta^2 d\theta \otimes d\theta~~~, 
\label{metric2}
\ee
where $x^{\mu}\equiv(x,y,\theta)$ with $\mu=1,2,3$.
The Dirac matrices on $H^2_r \times S^1_\beta$ are given in terms of flat 
Dirac matrices   by
\be
\gamma_1=  {r \over y} \sigma_1~,~~
\gamma_2=  {r \over y} \sigma_2~,~~
\gamma_3=  \beta \sigma_3 ~,
\label{VII.3}
\ee
while the spin connection  is
\be
\Gamma_{\mu}= {i\over 2y} \sigma_3 \delta_{1 \mu}~~~.
\label{spin2}
\ee
The v.e.v. of the $\psi$ fields being zero for this case too, we are left
with the gap equation
\be
0 = m \lim_{s\rightarrow 1}\biggl\{ \frac{1}{q(s)} -\zeta(s,m) 
\biggr\}~~~.                                
\label{gaph2s1}
\ee  
The equal-points heat kernel of the spin-${1\over 2}$ Laplacian on $H^2$
is \cite{VD}
\be
h_{H^2_r}(t; z= z')=\frac{r}{2(\pi t )^{3/2}} 
	\int_0^\infty dx~x~\coth x       
\exp\left\{ {-x^2r^2\over t}\right\}~~~, \label{hkerh2}
\ee
whereas the equal-points heat kernel of the scalar Laplacian on
the circle is given by Eq. \eqn{hkers1}. Substituting  in Eq. \eqn{mellin}, the
zeta-function reads
\beqa
\zeta(s,m)&=& { r\over 2\pi^{3/2}\beta \Gamma(s)} 
\int_0^\infty dt~t^{s-5/2}  
\Biggl\{ 
\int_0^\infty dx ~x~\coth x   \nn\\
&\times& \sum_{n=-\infty}^{\infty}
\exp \left[ -  {x^2r^2\over t} - 
\left( {4\pi^2\over \beta^2}\left(n+{1\over 2}\right)^2 + m^2
\right)t \right]\Biggr\}~.  
\label{zh2s1}
\eeqa
 Performing  the integral in $t$ and taking the 
limit $s\rightarrow 1$ the gap
equation \eqn{gaph2s1} becomes:
\beqa
m\left\{{ 1\over \pi\beta} 
\int_0^\infty {dx\over x} (x~\coth x - 1)
~\sum_{n=0}^{\infty} \exp\left[-2xr\sqrt{{4\pi^2\over \beta^2}
(n+{1\over 2})^2 +m^2}\right] \right. \nn\\
\left . - {m\over 4\pi} -
{1\over 2\pi\beta}\log(1+e^{-\beta m}) -{1\over q}\right\} = 0.
\label{ggaph2s1}
\eeqa
For details of the  calculation we refer to  appendix B.
 Let us discuss the solutions for $\beta$ finite first.
The first term in Eq. \eqn{ggaph2s1} is positive definite, hence, 
a non zero value of the mass
is possible in principle in the whole space of the parameters $r,\beta,
{1\over q}$, even at the critical coupling ${1\over q}= 0$ (in
\cite{MV} this point is missed).  $m=0$ is a solution as well. In
this limit the gap equation \eqn{ggaph2s1} becomes indeed
\be
(\lim_{m\rightarrow 0} m) \left\{ { 1\over 4\pi\beta} 
\int_{-\infty}^\infty {dx\over x} (x~\coth x - 1)\mbox{cosech} 
({2\pi r\over\beta} x)
- {1\over 2\pi\beta}\log2  -{1\over q}\right\} = 0 \label{gapm0}
\ee
where the factor  multiplying $m$ is finite. 
In facts by the study of the second derivative of the effective potential
(minus the first derivative of Eq. \eqn{ggaph2s1}), the value $m=0$ is seen to
be either a min or a max, depending on the values of the parameters,
$\beta, r, {1\over q}$. This is the second order phase transition point.
 Let us try a more quantitative analysis. 
To find the critical surface we proceed as in the positive curvature case. 
The transition being of second order, the mass gap, defined by
Eq. \eqn{ggaph2s1} up to a factor of $m$, smoothly disappears as we approach
criticality. Hence, performing the 
 limit $m\rightarrow 0$, we get
\be
  {2\pi\beta\over q}= {1\over 2} \int_{-\infty}^\infty {dx\over x}
(x~\coth x -1)\mbox{cosech} 
({2\pi r\over\beta} x)
- \log2  \label{h2s1crili}
\ee
This equation, which is the analogue of Eq. \eqn{s2s1crili} for the negative
curvature case, is numerically solved in the same way. 
Moreover, it can be solved 
analytically for specific values of the parameters. 
For $2\pi r=\beta$, the integral in Eq. \eqn{h2s1crili} simplifies 
yielding $\log 2$. Hence, when the coupling is zero (the flat space critical
value), we have
\be
{r_{cr}\over \beta_{cr}} = {1\over 2\pi }. \label{tcri}
\ee
When the coupling is positive, 
  \be
{r_{cr}\over \beta_{cr}} < {1\over 2\pi  }. \label{ttcri}
\ee  
whereas, for negative coupling
\be
{r_{cr}\over \beta_{cr}} > {1\over 2\pi }. \label{tttcri}
\ee  
 Differently from what happens in the positive curvature case, we
have a phase transition also for positive and zero values of the 
coupling. The latter result in particular,  tells us that the flat
space critical coupling, $1/q=0$,  which is a second order phase
transition point at zero curvature and temperature, maintains this
property when $r$ and  $ \beta$ meet the condition \eqn{tcri}.

In figure 2 we plot the temperature ${1/ \beta_{cr}}$
as a function of the inverse radius ${1/r_{cr}}$ for positive, negative, and 
zero values of the inverse coupling. For $1/q$ non-zero, the two radii 
are rescaled by the common factor $4 \pi/|q|$.The behaviour predicted by  
 Eqs. \eqn{tcri}-\eqn{tttcri}, can be verified to hold. 
 The zero curvature limit of Eq. \eqn{h2s1crili} may be performed yielding 
\be
{1 \over 2 \pi \beta} \log 2= -{1\over q}
\ee
which coincides with  Eq. \eqn{0curv} when the coupling constant is negative,
whereas it has no solution in the other cases.

Let us now  consider the zero temperature limit. Since the characteristics
are quite different, it deserves a careful analysis.
If we take the limit $\beta\rightarrow \infty$ of Eq. \eqn{gapm0}, we get 
a divergent result. This means that $m=0$ is never a solution of the gap
equation at zero temperature. In other words, the symmetry is always
broken, for any value of the coupling and of the curvature parameter (this
is to be compared with the positive curvature case, where at zero
temperature there is a curvature-induced phase transition).
The gap equation for this case is to be obtained by Eq. \eqn{ggaph2s1} taking 
 the limit $\beta\rightarrow\infty$, or, more directly, by the zeta
function for the space $H^2\times R$. This reads
\be
\zeta(s)_{H^2\times R}(s,m)= { r\over 4\pi^{2} \Gamma(s)} 
\int_0^\infty dt~t^{s-3}  
\int_0^\infty dx ~x~\coth x   
\exp \left( -  {x^2r^2\over t} -  m^2 t \right)~.  
\label{zh2R}
\ee
and it is obtained by replacing the
equal points heat
kernel of the scalar Laplacian on $S^1$ with the one on $R$, in
the equation \eqn{zh2s1}. Replacing Eq. \eqn{zh2R} in the gap equation 
\eqn{gaph2s1} 
and performing the integral in $t$ 
we easily arrive at  the gap equation 
\be
m\left\{{ m\over 2\pi^2} 
\int_0^\infty {dx\over x} (x~\coth x - 1) K_1(2xrm)
 - {m\over 4\pi}  -{1\over q}\right\} = 0.
\label{gaph2r}
\ee
 The integral on the LHS being positive for any value of $r$, and
divergent for $m=0$, this
equation yields   a non zero value of the mass whatever are the sign 
and the value of the coupling constant. 

The value of the mass gap is plotted in Fig. 3 as a function of the
curvature parameter $r$, for different values of the inverse coupling. 
At the critical coupling it takes the simple expression
\be
m(q_{cr})\simeq {1\over 4 r} .\label{masgap}
\ee 
This equation, which has the right behaviour for $r\rightarrow \infty$,
has an interesting property: it states that the mass gap at the critical
coupling is  half the mass gap which is found in \cite{GRV} for the
conformal sigma model on the same space.  
 We think that it is an
interesting result because it is a manifestation of bosonization, 
although  in
the large $N$ approximation (another warning is the fact that the result
\eqn{masgap} is
numerical,
whereas the result  in \cite{GRV} is analytic). 

We now consider the free energy density for the case of $\beta$ finite.
The calculation is  performed in  appendix B.
Taking the derivative of the zeta function \eqn{zh2s1},
 the free energy density \eqn{regfen} is seen 
to be
\beqa
{\cal W} (m)&=&{ 1\over 4\pi\beta r^2} 
\left(1 - r{d\over dr} \right) \int_0^\infty {dx\over x^2} ( 
{1\over x}
+ {x\over 3} -\coth x)
 \sum_{n=0}^{\infty}
 \exp\left(-2xr\sqrt{{4\pi^2\over \beta^2}
(n+{1\over 2})^2 +m^2}\right)\nn\\
&¬-&¬ {1\over 2\pi\beta^3 }\left( 1- \beta{\del\over\del\beta}\right)
\sum_1^\infty
{(-1)^n\over n^3} e^{-n\beta m} \nn \\
&+& {m\over48\pi r^2}   -
{m^3 \over 12\pi} + {1\over 48\pi\beta r^2} \log\left (1+  e^{-\beta
m}\right)
 -{m^2\over 2 q}.
\label{h2s1fe}
\eeqa
This expression, which looks complicated at a first sight, reproduces
correctly some known limits, as can be easily checked, like, for example, the
flat space limit \eqn{fren0}. In particular it allows
to calculate easily the free energy density on the critical surface.  
Upon taking the limit $m\rightarrow 0$ we arrive at
\beqa
{\cal W}_{H^2\times S^1}(0)&=&{1\over 8\pi\beta r^2 } 
 \int_0^\infty {dx \over x^2} \left[ 
{1\over x} +{x \over 3}- \coth x \right] \left[ 1+ {2\pi r\over\beta} x
\coth\left({
2\pi r x \over\beta}  
\right)\right] {1 \over \sinh (2 \pi r x/\beta) } \nn\\
&+ &  {3\zeta_R(3)\over 8 \pi \beta^3} 
+{\log 2 \over 48 \pi \beta r^2}   ~~~,
\eeqa
which yields the critical value of the energy density when $r$ , $\beta$,
and the coupling constant  meet 
the conditions \eqn{tcri}-\eqn{tttcri}. The analogous result reported in
\cite{MV} contains an error in the calculation. 
The limit of zero curvature yields the same expression as Eq. \eqn{0cu}.
The  zero temperature energy is to be  derived either from Eq. \eqn{h2s1fe} in
the
limit
$\beta\rightarrow \infty$, or directly from the zeta function
\eqn{zh2R}.  By means of the relation \eqn{dzeta}, which is true in this 
case as well, we find 
\be
{\cal W}_{H^2\times R} = -{m^2\over 4\pi^2 r } 
 \int_0^\infty {dx \over x} \left[\coth x 
-{1\over x} -{x \over 3} \right] K_2(2xrm) - {m^3\over 12\pi} + {m\over 48 
\pi r^2} -{m^2\over 2 q}.
\ee

\section{Conclusions}
Using the large $N$ approximation and the zeta-function regularization
to study the Gross Neveu model in 3 dimensions, we have found that it
describes a system undergoing a 
second order phase transition when considered on spaces of
constant curvature, at finite temperature. 
We have found the critical surface $f(r,\beta, q)= 0$, for de Sitter and 
anti-de Sitter spaces, where $r, \beta, q$
are respectively, the curvature parameter, the inverse temperature, and
the coupling constant. In both cases (positive and negative curvature) 
this turns out to be a critical line, 
$f(r/q,\beta/q)= 0$, the coupling constant appearing as a common scale
factor when finite. 
 In the case of positive curvature this is
represented  by 
Eq. \eqn{s2s1crili}, and  phase transitions are allowed only in the
negative region of the coupling constant. For the negative curvature 
case we have 
Eq. \eqn{h2s1crili} which describes phase transitions for an arbitrary
value of the coupling. In particular at the flat space critical coupling we 
have found an explicit solution, Eq. \eqn{tcri}, which states that the
product of temperature and curvature is constant on the critical line.
At zero temperature the symmetry is broken for any value of the curvature
and of the coupling. In this case it is possible to calculate the mass
gap. We have evaluated it at the flat space critical value of the 
coupling,  
Eq. \eqn{masgap}, and we
have found that this is half the mass gap of the conformal sigma model
when studied under the same assumptions.

\noindent \appendice

This appendix contains a proof of the results \eqn{lgaps2s1} and
\eqn{s2s1fe} of the subsection {\bf 2.1}.

Let us start with the definition of the zeta-function \eqn{zs2s1}
\be
\zeta(s,m)= 
{\beta^{2s-1}\over 2 \pi^2 r^2
\Gamma(s)} \int^{\infty}_0 dt~ t^{s-1}
\sum_{n=-\infty}^{\infty} \sum_{l=1}^{\infty} l \exp \Biggl\{-
\left[4\pi^2 \left( n+{1 \over 2}\right)^2  + l^2{\beta^2 \over r^2}
+m^2\beta^2 \right]t\Biggr\}.
 \label{zs2s1a}
\ee
To exchange the sum over $n$ with the integral in $t$ we use  the Poisson
sum formula for the heat kernel of the
spin-${1\over 2}$ Laplacian on the
circle,
\be
\sum_{-\infty}^{\infty}
\exp\left[-4\pi^2\left(n+{1\over 
2}\right)^2 t\right] = {1\over \sqrt{4\pi t}}\left[1+ 2  
\sum_{1}^{\infty} (-1)^n
\exp\left(-{n^2\over 4t}\right) 
\right]  \label{poisum}
\ee
 so that
\beqa
\zeta(s,m) &=& {\beta^{2s-1}\over 4 \pi^{3/2} r^2
\Gamma(s)} \left\{ \int^{\infty}_0 dt~ t^{s-3/2}
 \sum_{l=1}^{\infty} l \exp \Biggl\{-
\left[ {\beta^2\over r^2}l^2 +m^2\beta^2 \right]t\Biggr\} \right. 
\label{a+b} \\
&+&\left. 
2 \sum_{n=1}^{\infty} (-1)^n \int^{\infty}_0 dt~ t^{s-3/2}
 \sum_{l=1}^{\infty} l \exp \Biggl\{-
\left[ l^2{\beta^2 \over r^2} +m^2\beta^2 \right]t -{n^2 \over 4
t} \Biggr\} \right\}  = A+B \nn
\eeqa
The first integral, $A$, is performed by means of the Poisson sum formula
for the heat kernel of the spin-${1\over 2}$ Laplacian on the
sphere
\be
\sum_{l=1}^{\infty} l \exp \left\{-
\left[ l^2{\beta^2 \over r^2}\right]\right\} = {r^3\over \beta^3}
{t^{-3/2}\over 2\sqrt{\pi}} \int_{-\infty}^{\infty} dx \; x \cot x
\exp\left(-x^2{r^2\over\beta^2} \right). \label{lpoisum}
\ee
This is a specialization of the usual Poisson sum
formula.  A derivation may be found  for example in Appendix B of 
\cite{MV}.
This formula allows us to exchange the sum over $l$ with the integral in
$t$, upon extracting the spurious divergences which eventually show up. We
obtain
\beqa
A&=& {m^{2-s} r^{s-1} \over 2 \pi^{2} \Gamma(s)} 
\int^{\infty}_0 {\mbox P}\int_{0}^{\infty} dx ~ x^{s-1}\left(\cot x
-{1\over x}\right) K_{s-2}(2mxr) \nn\\
&+& {m^{-2s+3} \Gamma(s-{3\over 2})\over 8\pi^{3/2} \Gamma(s)} \label{a1}
\eeqa
where $K_\nu(x)$ is the modified Bessel function.
The second integral is easier to calculate. It yields
\be
B= {\beta^{s-1/2}2^{1/2-s}\over\pi^{3/2}r^2 \Gamma(s)}
\sum_{n,l=1}^{\infty} (-1)^n n^{s-1/2}{l\over({l^2\over r^2}
+m^2)^{s/2-1/4}} K_{s-1/2}\left(n\beta\sqrt{{l^2\over r^2} + m^2}\right) 
\label{b1}
\ee
where we have used \cite{prud}
\be
\int_0^\infty dt t^{\nu-1}e^{-at -b/t} = 2 ({b\over a})^{\nu/2}K_\nu
(2\sqrt{ab}),~~~a,b>0~. \label{pru}
\ee
The zeta-function is then rewritten as
\beqa
\zeta (s,m)&=& {m^{2-s} r^{s-1} \over 2 \pi^{2} \Gamma(s)} 
\int^{\infty}_0 {\mbox P}\int_{0}^{\infty} dx ~ x^{s-1}\left(\cot x
-{1\over x}\right) K_{s-2}(2mxr) \nn\\
&+& {m^{-2s+3} \Gamma(s-{3\over 2})\over 8\pi^{3/2} \Gamma(s)} 
+ {\beta^{s-1/2}2^{1/2-s}\over\pi^{3/2}r^2 \Gamma(s)}\nn\\
&\times& \sum_{n,l=1}^{\infty} (-1)^n n^{s-1/2}{l\over({l^2\over r^2}
+m^2)^{s/2-1/4}} K_{s-1/2}\left(n\beta\sqrt{{l^2\over r^2} + m^2}\right) 
\label{z1s2s1}
\eeqa
Substituting in \eqn{gaps2s1} and taking the limit $s\rightarrow 1$, we
get \eqn{lgaps2s1}.

To obtain the expression of the free energy density \eqn{s2s1fe}, 
we first
rewrite the zeta-function \eqn{zs2s1a} in the form
\beqa
\zeta(s,m)&=&  {m^{2-s} r^{s-1} \over 2 \pi^{2} \Gamma(s)} 
\int^{\infty}_0 {\mbox P}\int_{0}^{\infty} dx ~ x^{s-1}\left(\cot(x)
-{1\over x} + {x\over 3}\right) K_{s-2}(2mxr) \nn\\
&+& {\beta^{s-1/2} 2^{1/2-s}\over\pi^{3/2} r^2 \Gamma(s)}
\sum_{n,l=1}^{\infty} (-1)^n n^{s-1/2}{l\over({l^2\over r^2}
+m^2)^{s/2-1/4}} K_{s-1/2}\left(n\beta\sqrt{{l^2\over r^2} + m^2}\right)
\nn\\
&+& {m^{-2s+3} \Gamma(s-{3\over 2})\over 8\pi^{3/2} \Gamma(s)}
-{1\over 48 \pi^{3/2}} {\Gamma(s-{1\over 2}) \over \Gamma(s) }
\;. \label{z0s2s1}
\eeqa
This expression differs by \eqn{z1s2s1} by terms which cancel out
when $s\rightarrow 1$. It is obtained exactly in the same way as the
previous one. The 
extra-terms are needed to regularize the expression when
$s\rightarrow 0$. 
When taking the derivative of this expression with respect to $s$, 
the only contribution at $s=0$ is given by the terms which contain the
derivative of $1/\Gamma(s)$, all the others being zero (they are
multiplied by a common $1/\Gamma(s)$). To be
more precise,
\be
{d\over ds}\zeta(s)|_{s=0}=
\left[\Gamma(s)\zeta(s)\left(-{\psi(s)\over\Gamma(s)}\right)
\right]|_{s=0}
= \left(\Gamma(s)\zeta(s)\right)|_{s=0},
\label{dzeta}
\ee
where we have used 
\be
\left({d\over d s} {1\over \Gamma(s)}\right)_{s=0}=
-\left({\psi(s)\over\Gamma(s)}\right)_{s=0}=1.
\ee
Hence, replacing \eqn{dzeta} in the definition of the free
energy, \eqn{regfen}, the equation \eqn{s2s1fe} is obtained. 

\noindent \appendice

This appendix contains a proof of the equations \eqn{ggaph2s1} and
\eqn{h2s1fe} of subsection {\bf 2.2}.
Let us start form the zeta-function \eqn{zh2s1},
\beqa
\zeta(s,m)&=& { r\over 2\pi^{3/2}\beta \Gamma(s)} 
\int_0^\infty dt~t^{s-5/2}  
\Biggl\{ 
\int_0^\infty dx ~x~\coth x   \nn\\
&\times& \sum_{n=-\infty}^{\infty}
\exp \left[ -  {x^2r^2\over t} - 
\left( {4\pi^2\over \beta^2}\left(n+{1\over 2}\right)^2 + m^2
\right)t \right]\Biggr\}~.  
\label{zh2s1m}
\eeqa
This is rewritten as
\beqa
\zeta(s,m)&=& { r\over 2\pi^{3/2}\beta \Gamma(s)} 
\int_0^\infty dt~t^{s-5/2}  
\Biggl\{ 
\int_0^\infty dx (x~\coth x -1)   \nn\\
&\times& \sum_{n=-\infty}^{\infty}
\exp \left[ -  {x^2r^2\over t} - 
\left( {4\pi^2\over \beta^2}\left(n+{1\over 2}\right)^2 + m^2
\right)t \right]\Biggr\}~ \\
&+&
{ 1\over 4\pi\beta \Gamma(s)} 
\int_0^\infty dt~t^{s-2}  
\sum_{n=-\infty}^{\infty}
\exp \left[  - 
\left( {4\pi^2\over \beta^2}\left(n+{1\over 2}\right)^2 + m^2
\right)t \right] = C+ D.\nn
\eeqa
To evaluate the contribution D we use the Poisson sum
formula \eqn{poisum}. We have
\beqa
D&=&{1\over (4\pi)^{3/2} \Gamma(s)} \int_0^\infty dt~t^{s-5/2} e^{-m^2 t}
\nn\\
&+&{1\over 4\pi^{3/2} \Gamma(s)} \sum_1^\infty (-1)^n \int_0^\infty
dt~t^{s-5/2} \exp\left(-m^2 t -{n^2 \beta^2\over 4t }\right).
\eeqa
We then use the result \eqn{pru}  to perform  the last integral (the first is
just a $\Gamma$ function). To evaluate the contribution C we use the 
result \eqn{pru} to perform the integral in $t$. 
Summing up we obtain
\beqa
\zeta(s,m)&=& { 2 r\over \pi^{3/2}\beta \Gamma(s)} 
\int_0^\infty dx (x~\coth x - 1)\label{zetah2s1}\\
&\times& \sum_{n=0}^{\infty}{x r\over\sqrt{{4\pi^2\over \beta^2}
(n+{1\over 2})^2 +m^2}}^{s-3/2} K_{s-3/2}\left(2 x r\sqrt{{4\pi^2\over \beta^2}
(n+{1\over 2})^2 +m^2}\right)\nn\\
&+& {1\over(4\pi)^{3/2}}{\Gamma(s-3/2)\over\Gamma(s)} m ^{3-2s} +
{1\over2\pi^{3/2} \Gamma(s)}\sum_1^\infty (-1)^n \left({nb\over
2m}\right)^{s-3/2} K_{s-3/2}(n\beta m). \nn
\eeqa
Replacing this expression into the gap equation \eqn{gaph2s1} and taking
the limit $s\rightarrow 1$ we obtain \eqn{ggaph2s1}.

To obtain the expression of the free energy density \eqn{h2s1fe}, we
follow the same procedure  as in Appendix A.
 We first
rewrite the zeta function as
 \beqa
\zeta(s,m)&=& { 2 r\over \pi^{3/2}\beta \Gamma(s)} 
\int_0^\infty dx \left(x~\coth x - 1-{x^2\over 3}\right)
\label{zetenh2s1}\\
&\times& \sum_{n=0}^{\infty}\left({xr\over\sqrt{{4\pi^2\over \beta^2}
(n+{1\over 2})^2 +m^2}}\right)^{s-3/2}
K_{s-3/2}\left(2xr\sqrt{{4\pi^2\over \beta^2}
(n+{1\over 2})^2 +m^2}\right)\nn\\
&+& {1\over(4\pi)^{3/2}}{\Gamma(s-3/2)\over\Gamma(s)} m ^{3-2s} +
{1\over2\pi^{3/2} \Gamma(s)}\sum_1^\infty (-1)^n \left({nb\over
2n}\right)^{s-3/2} K_{s-3/2}(n\beta m) \nn\\
&+& {1\over 48\pi^{3/2}}{\Gamma(s-1/2)\over\Gamma(s)} m ^{1-2s} +
{1\over24\pi^{3/2} \Gamma(s)}\sum_1^\infty (-1)^n \left({nb\over
2m}\right)^{s-1/2} K_{s-1/2}(n\beta m) . \nn
 \eeqa
This expression differs from \eqn{zetah2s1} by terms which vanish in the
limit $s\rightarrow 1$. As in the calculation of the energy for the
positive curvature case, these terms have been introduced to perform the
limit $s\rightarrow 0$, without introducing divergences when exchanging
integrals and sums. 

Observing that the derivative of the zeta-function is given by
\eqn{dzeta} 
and considering the limit $s\rightarrow 0$, the wanted expression,
Eq. \eqn{h2s1fe}, is obtained.
 
\section*{Acknowledgements}
The author wishes to thank  M. Asorey and J. Cari$\tilde n$ena at the 
Department of Theoretical 
Physics of the University of Zaragoza for warm hospitality and 
financial support while this work was completed. Support from 
INFN is also acknowledged.

\newpage

\begin{figure}[t]
\epsfysize=9cm
\epsfxsize=15cm
\epsffile{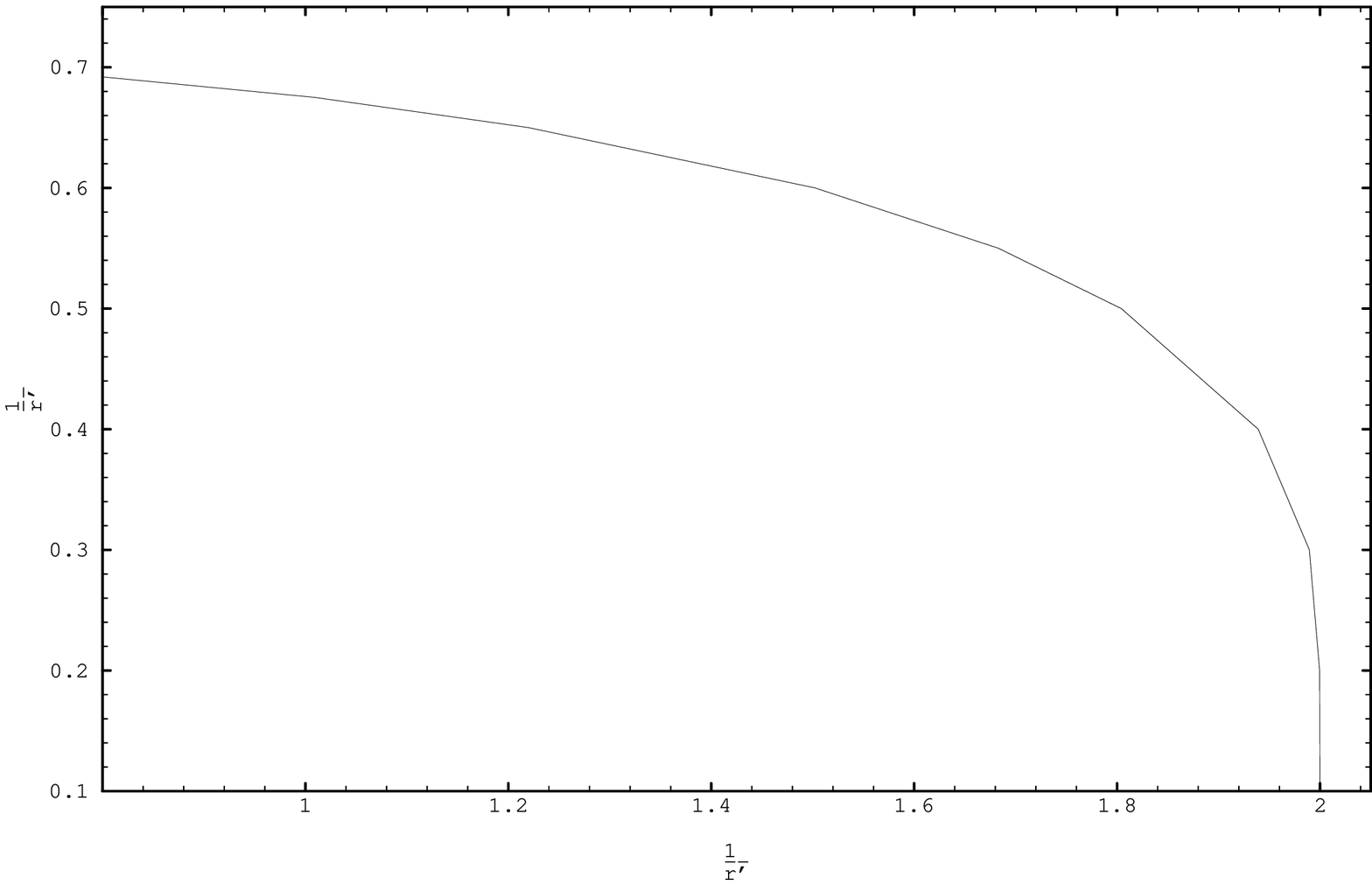}
\caption[Fig.1]
{The critical temperature  defined by \eqn{s2s1crili}, 
is plotted as a function of the  critical curvature parameter. They are 
both rescaled as  
 ${1\over b'}\equiv{1\over \beta m_0}$, ${1\over r'}
\equiv{1\over r m_0}$.}
\end{figure} 

\begin{figure}[t]
\epsfysize=9cm
\epsfxsize=15cm
\epsffile{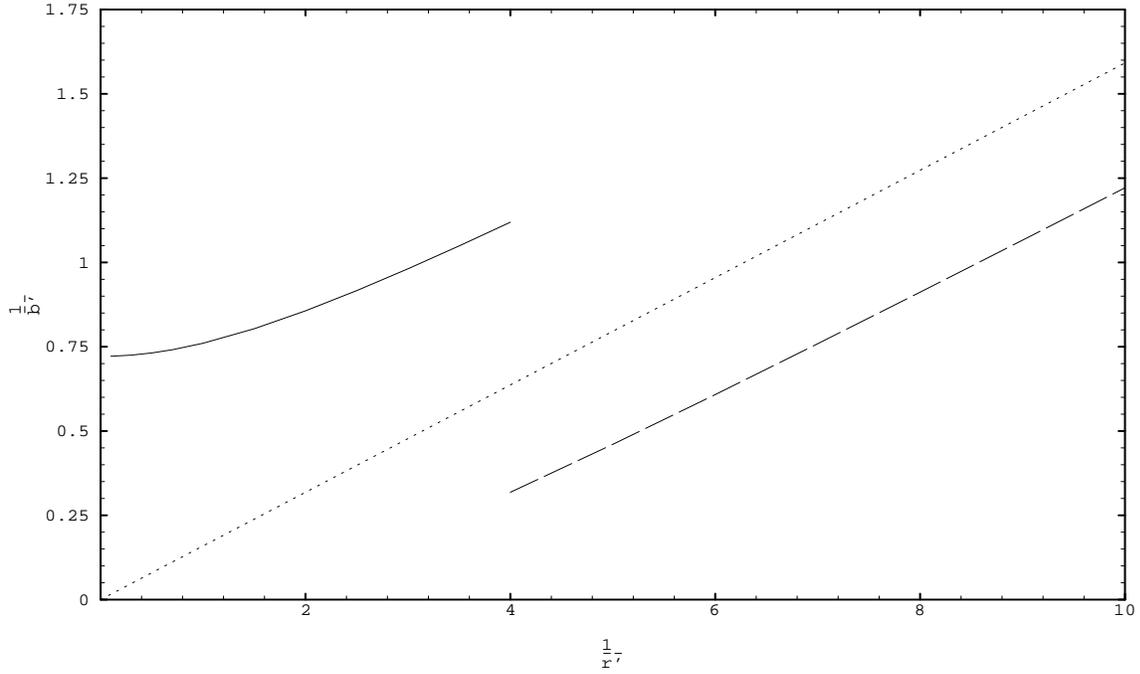}
\caption[Fig.2]
{The critical temperature   defined by \eqn{h2s1crili}, 
is plotted as a function of the  critical curvature parameter. 
The solid line corresponds to ${1\over q}<0$. The dashed line corresponds to 
${1\over q}>0$,
whereas the dotted line corresponds to ${1\over q}=0$. When ${1\over q}$ 
is negative or
positive, both the temperature and the inverse radius are rescaled to: 
${1\over b'}\equiv{4\pi\over |q|} {1\over \beta}$, 
${1\over r'}\equiv{4\pi\over |q|} {1\over r}$.}
\end{figure} 

\begin{figure}[t]
\epsfysize=9cm
\epsfxsize=14cm
\epsffile{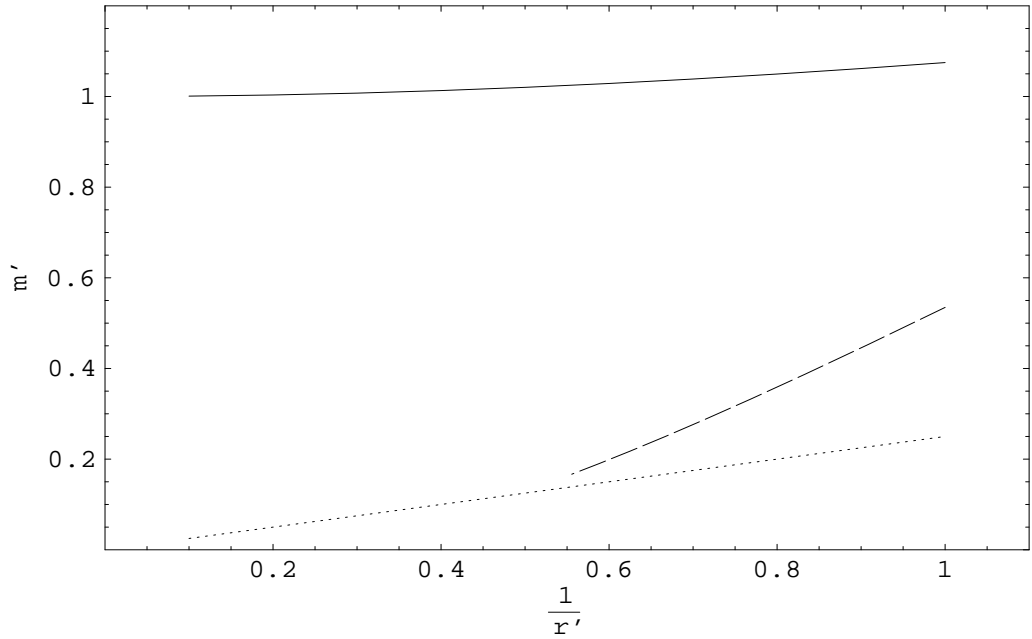}
\caption[Fig. 3] 
{The mass gap defined in Eq. \eqn{gaph2r}, is plotted as a function 
of the inverse radius.
The solid line corresponds to ${1\over q}<0$. The dashed line corresponds to 
${1\over q}>0$,
whereas the dotted line corresponds to ${1\over q}=0$. When ${1\over q}$ 
is negative or
positive, both the mass and the radius are rescaled to: 
$m'={4\pi\over |q|} m$, $r'={|q|\over 4\pi} r$.}
\end{figure} 
\end{document}